\documentclass[pra,onecolumn,aps,showpacs,footinbib,bibnotes]{revtex4-2}
\usepackage{graphicx}
\usepackage{physics,amsmath,amsfonts,amssymb,dsfont,amsthm}
\usepackage{mathtools}
\usepackage{multirow}
\usepackage[utf8]{inputenc}
\usepackage{comment}
\usepackage{subcaption}

\theoremstyle{plain}

\newcommand{\ie}{\textit{i}. \textit{e}.}

\usepackage{xcolor}
\definecolor{maroon}{RGB}{100,20,20}
\definecolor{dblue}{RGB}{20,20,100}
\usepackage[colorlinks=true,linkcolor=dblue,citecolor=blue,urlcolor=blue]{hyperref}
\begin{document}
\title{Security in a prepare-and-measure quantum key distribution protocol when the receiver uses weak values to guess the sender's bits}
\author{Rajendra Singh Bhati}
\email{rsbhati@cft.edu.pl}
\affiliation{Center for Theoretical Physics, Polish Academy of Sciences, Aleja Lotników 32/46, 02-668 Warsaw, Poland}

\begin{abstract}
The weak values and weak measurement formalism were initially limited to pure states which was later extended to mixed states,
leading to intriguing applications
in quantum information processing tasks. 
Weak values are considered to be abstract properties of
systems describing a complete picture between
successive measurements in the two-state vector formalism (TSVF).
The remarkable
achievements of the weak value formalism in experimental quantum mechanics have
persuaded most of quantum physicists that it is impeccable. However,
we explore a scenario where the formalism of weak values for mixed states is
employed in a quantum communication protocol but discover that it generates
inaccurate outcomes. This reinforces our previous conclusion that the weak values
may not be elements of the reality of weak measurements, contrary to what the
proponents of weak values proposed.
\end{abstract} 

\maketitle

\section{Introduction}

Weak values, together with the two-state vector formalism (TSVF)~\cite{PhysRevA.41.11,Aharonov_1991,PhysRevA.96.032114}, provide a framework for describing the physical properties of pre- and post-selected (PPS) quantum systems. The concept of weak values relies on the intriguing phenomenon of weak measurement, which allows experimenters to extract information from quantum systems while introducing only minimal disturbance. In such measurements, the readout corresponds to the weak value of the observable being measured, given that the system is post-selected after the measurement interaction. Although the emergence of weak values in the post-processing of the pointer state is often attributed to the neglect of higher-order perturbations in the system–pointer interaction, proponents of TSVF interpret this phenomenon as evidence of a deeper, time-symmetric structure in quantum mechanics—one that is dictated by boundary conditions in time imposed by both past and future measurement outcomes.

Despite ongoing scholarly debate surrounding their interpretation and foundational significance~\cite{vaidman1,PhysRevA.96.022126,BHATI2022127955,REZNIK2023128782,Alonso2015,PhysRevA.94.032115,SOKOLOVSKI2017227}, the theory of weak measurement and weak values has proven to be a powerful tool in a wide range of quantum information processing tasks. These include applications such as quantum process and state tomography~\cite{Lundeen2011,PhysRevLett.108.070402,Kim2018}, ultrasensitive quantum measurements through weak-value-based signal amplification~\cite{Hallaji2017,PhysRevLett.107.133603,PhysRevLett.112.200401,PhysRevLett.113.030401}, and investigations into fundamental problems such as the reconstruction of Bohmian trajectories in the double-slit experiment~\cite{Mahlere1501466}, the Hardy paradox~\cite{MOLMER2001151}, superluminal and slow-light phenomena~\cite{PhysRevLett.92.043601, PhysRevLett.93.203902}, quantum tunneling times~\cite{PhysRevLett.74.2405,PhysRevA.52.32}, and many others.
The weak values and weak measurement formalism were initially limited to pure 
states~\cite{PhysRevLett.60.1351,PhysRevLett.66.1107,PhysRevA.41.11}.
However, it was later extended to mixed states~\cite{PhysRevA.65.032111,PhysRevA.89.012121,PhysRevLett.114.090403,PhysRevA.96.032114},
leading to intriguing applications
in quantum information processing tasks~\cite{PhysRevLett.108.070402, Kim2018}.

In this work, we examine the potential use of generalized weak values in quantum key distribution (QKD) and identify a flaw that, if overlooked, may lead to misleading conclusions about quantum security. Moreover, we propose a quantum state discrimination (QSD) scheme that can be incorporated into a prepare-and-measure QKD protocol to reduce the quantum bit error rate (QBER). Our analysis shows that a straightforward application of generalized weak values in this context can yield a protocol that appears secure within the weak-value formalism but is, in fact, not. This false sense of security arises from the weak measurement approximation, wherein higher-order terms in the interaction strength are neglected.

The problem of quantum state discrimination is central to quantum communications~\cite{PERES198819,Chefles2000,Bae_2015}. In a typical protocol, a sender (Alice) transmits a quantum system prepared in one of several possible states to a receiver (Bob), or equivalently, steers Bob’s system via shared quantum correlations. Bob’s goal is to identify the transmitted state with minimal error using only local resources. However, the communication channel is often noisy, allowing an eavesdropper (Eve) to gain partial information about the transmitted states~\cite{RevModPhys.81.1301,renner2006security,PhysRevA.72.012332}. The same noise contributes to the QBER observed by Bob, thereby reducing the secure key rate. A QKD protocol remains secure only if the mutual information shared between Alice and Bob exceeds that accessible to Eve, modeled through her quantum memory~\cite{Devetak2005,renner2006security,PhysRevA.72.012332}. Security can thus be enhanced either by restricting Eve’s information gain or by improving Alice and Bob’s correlations—particularly through better state discrimination on Bob’s side.

The success probability in minimum-error discrimination (MED) strategies is fundamentally limited by the Helstrom–Holevo bound~\cite{helstrom1969quantum,holevo2011probabilistic}. Interestingly, weak measurements and weak values can be leveraged to achieve improved discrimination performance. Using the formalism of generalized weak values, we design a QSD scheme that reduces the QBER and enhances the correlations between Alice and Bob, thereby improving the noise tolerance of the protocol. Before presenting this scheme, we derive a general expression for generalized weak values from first principles using the TSVF and demonstrate that it constitutes a legitimate extension of the original weak value concept, initially formulated for pure states. However, a careful security analysis performed without invoking the weak measurement approximation reveals that this approach offers no real advantage in the secure key rate. We further examine the origin of this discrepancy and discuss its implications for the use of weak-value-based methods in quantum information processing.

This article is organized as follows. Section~\ref{sec_3.2} presents a concise derivation of generalized weak values. In Sec.~\ref{sec_3.3}, we introduce a scheme for state discrimination based on weak values, and Sec.~\ref{sec_3.4} describes a quantum key distribution (QKD) protocol that employs this technique. The proposed protocol is a modification of the six-state protocol~\cite{PhysRevLett.81.3018}, in which Bob uses the weak-value-based state discrimination strategy to infer Alice’s bit. Section~\ref{sec_3.5} defines the security criteria for the protocol, while Sec.~\ref{sec_3.6} analyzes its security under the weak measurement approximation (WMA), where higher-order terms in the interaction strength are neglected. We emphasize that the WMA ensures that the pointer-state displacements are linearly proportional to the weak values. Thus, adopting the WMA implicitly assumes that weak values faithfully represent elements of reality in weak measurements. We derive the joint probability distributions for Alice and Bob and estimate the eavesdropper’s quantum memory, assuming a depolarizing quantum communication channel. Our results show that incorporating weak values improves the noise tolerance of the six-state protocol (SSP). Section~\ref{sec_3.7} presents a security analysis of the protocol without invoking the WMA and demonstrates that, when all orders of the interaction strength are retained in the key-rate calculation, the protocol offers no advantage over the original six-state protocol. Finally, Sec.~\ref{sec_3.8} summarizes the main results and discusses their implications.

\section{Weak values and weak measurements}\label{sec_3.2}

The weak value of an observable $\boldsymbol{A}$ for a system pre-selected in state $\ket{\psi}$
and post-selected in state $\ket{\phi}$ is defined as~\cite{PhysRevLett.60.1351}
\begin{equation}
	\label{wv}
	\langle \boldsymbol{A}\rangle_w=\frac{\bra{\phi}\boldsymbol{A}\ket{\psi}}
	{\bra{\phi}\ket{\psi}}.
\end{equation}
In a weak measurement scenario involving a pre- and post-selected (PPS) system, the displacement of the pointer state is directly proportional to the weak value of the measured observable.
Consider a pointer $P$ initially prepared in a Gaussian wave packet centered at the origin in the position basis:
\begin{equation}
\label{pointer}
	\xi(x)=(2\pi\delta^2)^{-1/4}\exp(-{x^2}/{4\delta^2}),
\end{equation}
where $\delta$ characterizes the width of the packet. The pointer interacts with a system 
$S$, initially prepared in state 
$\ket{\psi}$, through the unitary evolution $U_{SP}=\exp(-i\gamma\boldsymbol{A}\otimes\hat{p})$
generated by a von Neumann type interaction Hamiltonian $H_{int} = g(t)\boldsymbol{A}\otimes\hat{p}$ where $\gamma = \int_{0}^{\infty}g(t)dt \ll 1$ is the interaction strength,
$\boldsymbol{A}$ is the system observable,
and $\hat{p}$ is the momentum operator of the pointer.
After post-selecting the system in state $\ket{\phi}$, the pointer's wavefunction in the position basis becomes 
\begin{equation}\label{e2.2}
	\xi^\prime(x)= (2\pi\delta^2)^{-1/4}e^{i\gamma\Im{\langle \boldsymbol{A}\rangle_w}x}\exp\left(-\frac{\left(x-\gamma\Re{\langle \boldsymbol{A}\rangle_w}\right)^2}{4\delta^2}\right).
\end{equation}
Here, we have assumed $\gamma^2\approx 0$ and retained only first-order terms
in interaction strength, which characterizes weak measurements. The real and imaginary parts of the weak value $\langle \boldsymbol{A}\rangle_w$ can be measured directly by measuring the position and momentum shifts of the pointer.

Let us now revisit the generalization of Eq.~\eqref{wv} ot mixed states, as presented by ref.~\cite{PhysRevA.96.032114}.
Instead of pre-selection in the pure state $\ket{\psi}$, consider
the case where the system is prepared in a mixed state $\rho=\sum_{i}p_i\ketbra{\psi_i}$
and post-selected in the state $\ket{\phi}$. A purification of $\rho$,
denoted by $\ket{\Psi}$, can be
written by introducing an ancillary system with a set of orthogonal states $\{\ket{e_i}\}$, as
\begin{equation}
	\label{Psi_puri}
	\ket{\Psi} = \sum_{i}\sqrt{p_i}\ket{\psi_i}\otimes\ket{e_i}.
\end{equation}
The preparation of the system in $\rho$ is operationally equivalent to the pre-selection
in the composite state $\ket{\Psi}$ of the system and the ancilla.
The post-selection of the system in $\ket{\phi}$ is equivalent to
performing a post-selection measurement $\mathcal{M}_{post}$, given by
\begin{equation}
	\mathcal{M}_{post} = \{\ketbra{\phi}\otimes\mathds{1}, \mathds{1}\otimes\mathds{1}-\ketbra{\phi}\otimes\mathds{1}\},
\end{equation}
on the combined system and selecting outcomes corresponding to the
projection $\ketbra{\phi}\otimes\mathds{1}$. The joint state after the
post-selection becomes
\begin{equation}
	\label{Phi_puri}
	\ket{\Phi} = N\ket{\phi}\otimes\sum_i\sqrt{p_i}\braket{\phi}{\psi_i}\ket{e_i},
\end{equation}
where $N$ is the normalization factor. Since the system and the ancilla are jointly pre-and post-selected in pure states, we can apply Eq.~\eqref{wv} to derive the expression for the weak value of the local observable $\boldsymbol{A}$ as
\begin{equation}
	\langle \boldsymbol{A}\rangle_w=\frac{\bra{\Phi}\boldsymbol{A\otimes\mathds{1}}\ket{\Psi}}
	{\bra{\Phi}\ket{\Psi}}.
\end{equation}
Using Eqs.~\eqref{Psi_puri} and \eqref{Phi_puri}, we get
\begin{equation}
\label{gen_wv}
	\begin{aligned}
		\langle\boldsymbol{A}\rangle_w &= \frac{\bra{\phi}\otimes\sum_i\sqrt{p_i}\braket{\psi_i}{\phi}\bra{e_i}\sum_j\sqrt{p_j}\boldsymbol{A}\ket{\psi_j}\otimes\ket{e_j}}
		{\bra{\phi}\otimes\sum_i\sqrt{p_i}\braket{\psi_i}{\phi}\bra{e_i}\sum_j\sqrt{p_j}\ket{\psi_j}\otimes\ket{e_j}} \\
		& = \frac{\sum_i p_i\braket{\psi_i}{\phi}\bra{\phi}\boldsymbol{A}\ket{\psi_i}}
		{\sum_i p_i\braket{\psi_i}{\phi}\bra{\phi}\ket{\psi_i}} \\
		& = \frac{\bra{\phi}\boldsymbol{A}\rho\ket{\phi}}
		{\bra{\phi}\rho\ket{\phi}}.
	\end{aligned}
\end{equation}
An interesting aspect of this derivation is that we have not imposed any specific assumptions on the pointer state or the weak measurement itself. Instead, we have relied solely on the TSVF, which asserts that the physical properties of a system between two successive measurements are represented by Eq.~\eqref{wv}. Consequently, Eq.~\eqref{gen_wv} serves as a natural and legitimate generalization of Eq.~\eqref{wv}, and all the implications of the TSVF extend to mixed states as well.

\section{State discrimination using weak values}\label{sec_3.3}

There are two main approaches for state discrimination: (1) minimum error discrimination (MED),
where states are distinguished with a nonzero error, and (2) unambiguous discrimination (UD)
in which the setup can distinguish input states with zero error, but can sometimes give
inconclusive answers~\cite{Chefles2000,Bae_2015}.
There can also be a mixture of these two strategies such that
the setup discriminates input states with nonzero error and gives inconclusive
answers with nonzero probability. Such a strategy can achieve an error probability
below the Helstrom-Holevo bound. 

Let us now consider an example where Bob is given a task to distinguish between two
Gaussian wavefunctions prepared with equal a prior probability,
\begin{equation}
	\begin{aligned}
		\psi_\pm(x) & = (2\pi\delta^2)^{-1/4}\exp(-\frac{(x\mp\epsilon)^2}{4\delta^2}) \\
	\end{aligned}
\end{equation}
The minimum error in MED for uniform a prior probability is given by
~\cite{helstrom1969quantum,holevo2011probabilistic,Bae_2015}
\begin{equation}
	P_{err} = \frac{1}{2}\left(1-\sqrt{1-|\braket{\psi_+}{\psi_-}|^2}\right)
\end{equation}
Since, $\braket{\psi_+}{\psi_-} = \int_{-\infty}^{\infty}\psi_+^{\ast}(x)\psi_-(x)dx=\exp(-\epsilon^2/2\delta^2)$, we have
\begin{equation}
	P_{err} = \frac{1}{2}\left(1-\sqrt{1-\exp(-\epsilon^2/\delta^2)}\right)
\end{equation}
Further, consider that the states given to Bob are very close to each other \ie~$\epsilon/\delta\ll 1$.
In this case, $P_{err}\approx\frac{1}{2}(1-\epsilon/\delta)$ meaning
Bob can only discriminate the given
states with the probability of order $\epsilon/\delta\ll 1$ using the MED strategy.
Let us now introduce a scheme to distinguish states with higher success probability,
but with a cost of inconclusive results. Bob measures the particle
in the position basis $x$.
If the particle is found at $x=\alpha$, the state is considered to be
$\ket{\psi_+}$, and if it is found at $x= -\alpha$,
the state is guessed to be $\ket{\psi_-}$ where $\alpha>0$.
The result is inconclusive if the particle is found in any other place.
Bob's action can be modeled mathematically by a measurement setting
$\mathcal{M}\equiv\{\Pi_+,\Pi_-,\Pi_?\}$ acting on the particle where 
$\Pi_+=\ketbra{\alpha}$,
$\Pi_-=\ketbra{-\alpha}$, and 
$\Pi_?=\mathds{1}-\Pi_+-\Pi_-$. Outcomes corresponding to
$\Pi_+$, $\Pi_-$, and $\Pi_?$ correspond to $\ket{\psi_+}$, $\ket{\psi_-}$, and
inconclusive results, respectively. The probability of incorrect identification of
the state conditioned on conclusive results can be evaluated as
\begin{equation}
	\begin{aligned}
		P_{err} & = \frac{\bra{\psi_+}\Pi_-\ket{\psi_+}+\bra{\psi_-}\Pi_+\ket{\psi_-}}{\bra{\psi_+}\Pi_-\ket{\psi_+}+\bra{\psi_-}\Pi_+\ket{\psi_-}+\bra{\psi_-}\Pi_-\ket{\psi_-}+\bra{\psi_+}\Pi_+\ket{\psi_+}} \\
		& = \frac{\exp(-(\alpha+\epsilon)^2/2\delta^2)}{\exp(-(\alpha+\epsilon)^2/2\delta^2)+\exp(-(\alpha-\epsilon)^2/2\delta^2)} \\
		& = \frac{1}{1+\exp(\frac{2\alpha\epsilon}{\delta^2})}
	\end{aligned}
\end{equation}

\begin{figure}
	\begin{center}
		\includegraphics[scale=0.6]{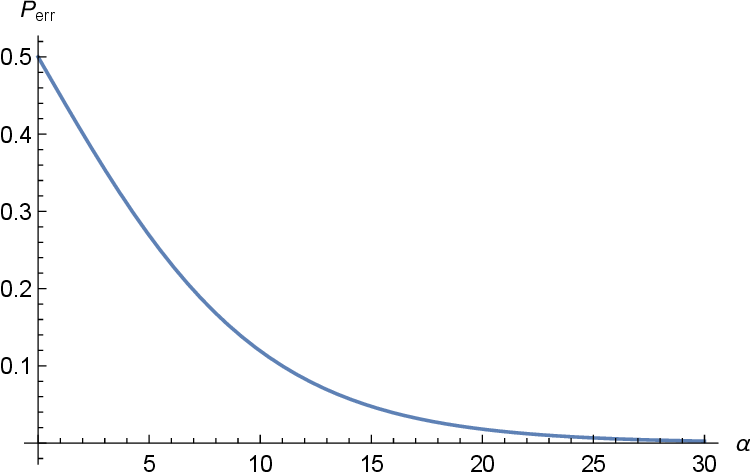}
		\caption{$P_{err}$ is plotted as a function of $\alpha$ for $\epsilon/\delta^2=0.1$
		}
		\label{fig3.1}
	\end{center}
\end{figure} 

As we can see in Figure~\ref{fig3.1}, $P_{err}$ decreases as $\alpha$ is increased
for constant $\epsilon/\delta^2$. In fact, it is possible to achieve an arbitrary low
error in state discrimination for given $\ket{\psi_+}$ and $\ket{\psi_-}$ but at a cost
of increased probability of inconclusive results.

Applied with weak measurements, the above strategy can be used to discriminate
states in Hilbert spaces of discrete dimensions. Suppose Bob is asked to discriminate
between two states $\ket{\phi_1}$ and $\ket{\phi_2}$ in a discrete dimensional space. Bob performs weak measurement of a carefully chosen observable
$\mathbf{A}$ of the given system using a pointer state prepared in the Gaussian state given by Eq.~\eqref{pointer} followed by post-selection
in a state $\ket{\phi}$. The pointer state transforms into
\begin{equation}
	\xi_i(x)=(2\pi\delta^2)^{-1/4}e^{i\gamma\Im{\langle \boldsymbol{A}\rangle^w_i}x}\exp\left(-\frac{\left(x-\gamma\Re{\langle \boldsymbol{A}\rangle^w_i}\right)^2}{4\delta^2}\right)
\end{equation}
where $i\in\{1,2\}$, $\gamma\ll 1$ is the interaction strength and
$\langle \boldsymbol{A}\rangle^w_i$ is the corresponding weak value given by
\begin{equation}
	\langle \boldsymbol{A}\rangle^w_i=\frac{\bra{\phi}\mathbf{A}\ket{\phi_i}}{\braket{\phi}{\phi_i}}
\end{equation}
It is easy to verify that Bob can always choose $\mathbf{A}$ and $\ket{\phi}$ in such a
manner that $\Re{\langle \boldsymbol{A}\rangle^w_1}=\beta$ and
$\Re{\langle \boldsymbol{A}\rangle^w_2}=-\beta$ for some $\beta\geq0$.
Bob's action can be modeled by a quantum map $\mathcal{B}(\cdot)$ that
transforms $\ket{\phi_i}$ into $\xi_i(x)$ \ie~$\mathcal{B}(\ket{\phi_i})=\xi_i(x)$.
Bob can now use the state discrimination strategy described above to discriminate between
$\xi_1(x)$ and $\xi_2(x)$ which is equivalent to discriminating $\ket{\phi_1}$ and
$\ket{\phi_2}$.
The use of weak values makes discrimination of arbitrary mixed states apparently plausible,
which is otherwise a non-trivial and mathematically difficult problem.
Suppose Bob is given a copy of two of the possible mixed states $\rho_1$ and $\rho_2$.
Similar to the pure-state case,
Bob can always find a suitable post-selection state and an observable $\mathbf{A}$
such that the pointer state $\xi(x)$ transforms to the desired $\xi_i(x)$. As we will see, the above strategy can be readily deployed in QKD protocols to improve the noise tolerance. However, as briefly mentioned in the introduction, it turns out to be flawed, and the reason is deeply rooted in the non-trivial connection between weak values of the mixed states and the pointer displacement in the weak measurement.

\section{QKD Protocol using weak values}\label{sec_3.4}

In a prepare-and-measure QKD protocol, Alice prepares a system in any of two pure
states say $\ket{0}$ and $\ket{1}$ or in $\ket{+}$ and $\ket{-}$ with equal probability
(as in BB84 protocol~\cite{BENNETT20147}),
and sends it to Bob. Assuming the channel to be depolarizing,
the sent state $\ket{\psi}\in\{\ket{0},\ket{1},\ket{+},\ket{-}\}$ transforms to
$\rho_{\psi}=(1-2\eta)\ketbra{\psi}+\eta\mathds{1}$, where $\eta\in[0,1/2]$ is the channel noise.
After guessing the correct basis, Bob applies a strategy to discriminate
between $\rho_{0}$ and $\rho_{1}$ (or between $\rho_{+}$ and $\rho_{-}$)
for raw key generation. In BB84,
Bob just measures the system in a correctly guessed preparation basis and generates
the key bit with a quantum bit error rate (QBER) equal to $\eta$.
Corresponding to every QKD protocol, there is maximally tolerated channel noise
$\eta_{tol}$ above which the protocol is considered to be insecure. The noise tolerance
of BB84 against collective attack is $\approx 11\%$, while the six-state 
protocol~\cite{PhysRevLett.81.3018} has a tolerance of $\approx 12.62\%$
\cite{PhysRevA.72.012332,RevModPhys.81.1301}.

In this chapter, first, we present a QKD protocol where Bob (the receiver) applies
the above-presented quantum state discrimination strategy using weak values for mixed states.
Assuming Eq.~\eqref{gen_wv} to be a valid expression for the weak values for
mixed states, and assuming the first-order approximation of weak measurements, we show that
such a QKD protocol can guarantee a secure key rate at an arbitrary high level of eavesdropping
\ie~at an arbitrary high $\eta_{tol}$. We present an information theocratic security proof
of the protocol against collective attacks while assuming the weak measurement
approximation (WMA) in which higher order terms in system-pointer interaction
unitary are neglected. WMA is at the center of weak measurement methodology
and has been validated by various experimental demonstrations~\cite{PhysRevLett.66.1107,
	PhysRevLett.94.220405,Lundeen2011,Kim2018}. Moreover, WMA has
played an important role in studies of various quantum paradoxes and phenomena
~\cite{PhysRevLett.91.180402,PhysRevLett.102.013902,PhysRevLett.92.043601,PhysRevLett.100.026804,
	Kocsis1170,PhysRevLett.111.240402,Denkmayr2014}.
We then re-analyze the security of the protocol without assuming WMA
\ie~retaining all terms in system-pointer interaction unitary. We find that
the protocol does not show tolerance against arbitrary high noise levels as
it appears in WMA analysis. Furthermore, it is observed that the noise tolerance is
in fact not better than BB84 or six-state protocols. Our results teach us non-trivial
aspects of WMA and weak values for mixed states. Contrary to what it is generally understood,
the use of weak values and weak measurements can sometimes mislead into completely wrong conclusions
and predictions.

Alice prepares an entangled qubit pair in state
$\ket{\Phi^{+}}=\frac{1}{\sqrt{2}}\left(\ket{00}+\ket{11}\right)$ and sends one
of the qubits to Bob via a quantum channel $\mathcal{E}(\cdot)$ while keeping
the other in her lab protected from any adversarial access. This step is repeated
$N$ number of times, where $N$ is asymptotically large. For simplicity, we
assume both parties have quantum memories and measurements can be postponed
to the end of the state sharing step. The protocol can easily be generalized to
memoryless scenarios as well.

Both parties then, agreeing over an authenticated classical communication (ACC),
divide the shared pairs into two parts where one is used for
parameter estimation and the second for raw key generation.
The choice of whether a pair is used for parameter estimation or key generation
is completely random and made after the completion of the successful sharing of systems.

Alice and Bob then use measurement settings of the six-state protocol to estimate the
channel noise as a (set of) parameter(s). More specifically, they randomly measure Pauli
operators $\sigma_x$, $\sigma_y$, and $\sigma_z$ and estimate errors $\varepsilon_x$,
$\varepsilon_y$, and $\varepsilon_z$, where $\varepsilon_i=P(a_i\neq b_i)$
is the probability of getting different outcomes when both parties measure the
same operator $\sigma_i$, $\forall i\in\{x,y,z\}$. For depolarizing channels,
$\varepsilon_x=\varepsilon_y=\varepsilon_z=\eta$ is the measure of channel noise.
If $\eta\geq\eta_{tol}$, for some $0\leq\eta_{tol}\leq 1/2$, they
abort the protocol, else they continue to raw key generation from the remaining
set of pairs.

Alice and Bob then execute the following steps to generate their raw keys
$X$ and $Y$, respectively, from the remaining set of pairs:

\begin{itemize}
	
	\item [1:] Bob prepares an ancillary system, we call it pointer here, in state $\ket{\xi}$
	specified by a Gaussian wave function 
	$\xi(x)=(2\pi\delta^2)^{-1/4}\exp(-x^2/4\delta^2)$ in the position basis.
	He then applies the unitary
	$U_{BP}=\exp(-i\gamma\sigma_z\otimes\hat{p})$ on the
	combined state of his qubit and the pointer such that $\gamma^2/\delta^2\ll 1$
	where $\hat{p}$ is the momentum operator of the pointer. In other other words, he perform weak measurement of $\sigma_z$ on his part of the shared Bell pairs.
	
	\item [2:] Alice performs measurement of the observable $\sigma_z$ on
	her qubit and records binary outcomes as
	$0$ and $1$ corresponding to eigenvalues $+1$ and $-1$, respectively.
	
	\item [3:] Bob then post-selects his qubit in the state $\ket{+}=\frac{1}{\sqrt{2}}\left(\ket{0}+\ket{1}\right)$.
	The rest of the rounds, \ie~corresponding to Bob's outcome 
	$\ket{-}=\frac{1}{\sqrt{2}}\left(\ket{0}-\ket{1}\right)$ in post-selection measurement,
	are discarded after agreeing over ACC.
	
	\item [4:] Thereafter, Bob performs measurement
	$\mathcal{M}\equiv\{\Pi_0,\Pi_1,\Pi_?\}$ on pointer where 
	$\Pi_0=\ketbra{\alpha}$,
	$\Pi_1=\ketbra{-\alpha}$, and 
	$\Pi_?=\mathds{1}-\Pi_0-\Pi_1$ for some $\alpha\geq0$.
	Rounds corresponding to Bob's outcome $\Pi_?$ are discarded after
	agreeing over ACC. Bob stores outcomes corresponding to $\Pi_0$ and $\Pi_1$
	as $0$ and $1$, respectively, and keeps them secret and protected from any adversarial
	access. This is Bob's raw key.
	
\end{itemize}

Alice and Bob now have partially secure and non-identical bit strings $X$ and $Y$
(raw keys), respectively, of equal length.
They then proceed to perform classical error correction (EC) and privacy amplification
(PA) on their raw keys to extract fully secure and completely identical keys.

\section{Security definition}\label{sec_3.5}
We consider security against collective attacks where the
same measurement strategy is applied on independent and identically distributed
(i.i.d.) quantum states and devices during every round of the protocol. Similarly,
Eve can also extract information from the quantum channel by interacting with shared
systems identically and independently in all rounds. Eve is always allowed to have quantum memory
and can postpone her measurements to the end of classical post-processing \ie~
EC and PA.

Let $\mathcal{H}_A$, $\mathcal{H}_B$, $\mathcal{H}_E$, and $\mathcal{H}_P$ be
Hilbert spaces of Alice's system, Bob's system, Eve's quantum memory, and Bob's pointer,
respectively. In each round, Alice and Bob share a bipartite state $\rho_{AB}=\mathcal{E}
(\ketbra{\Phi^+})$. Any noise introduced by channel $\mathcal{E}(\cdot)$ is attributed to
Eve's attempt of eavesdropping and thus the purification of $\rho_{AB}$
is described by a tripartite state $\ket{\Psi}_{ABE}$ distributed
among Alice, Bob, and Eve. The combined state, including Bob's pointer, can be
expressed (with respect to Bell basis in $\mathcal{H}_A\otimes\mathcal{H}_B$) as

\begin{equation}
	\ket{\Psi}_{ABEP}=\sum_{i=1}^4\sqrt{\lambda_i}\ket{\Phi_i}_{AB}\otimes\ket{\nu_i}_{E}\otimes\ket{\xi}_P
\end{equation}

where $\ket{\Phi_1}_{AB},\ket{\Phi_2}_{AB},\ket{\Phi_3}_{AB},\ket{\Phi_4}_{AB}$ are Bell 
states ${\ket{\Phi^+}},{\ket{\Phi^-}},{\ket{\Psi^+}},$ and ${\ket{\Psi^-}}$,
respectively, in $\mathcal{H}_A\otimes\mathcal{H}_B$ and $\{\ket{\nu_i}\}$ denotes a set of
orthogonal states forming a basis in Eve's state space $\mathcal{H}_E$.

Suppose that Alice and Bob prepare a bipartite system in the state $\ket{\Phi_i}$
and post-select in $\ket{\psi^a}=\ket{a}\otimes\ket{+}$ where $a\in\{0,1\}$,
after weak measurement of
the observable $\boldsymbol{\sigma}=\mathds{1}\otimes\sigma_z$ using interaction unitary $U_{BP}$.
This generates a translation in the pointer state
proportional to the weak value 
\begin{equation}
	\langle \boldsymbol{\sigma}^a_i\rangle_w=
	\frac{\bra{\psi^a}\boldsymbol{\sigma}\ket{\Phi_{i}}}{\bra{\psi^a}\ket{\Phi_{i}}}.
\end{equation}
If the initial wave function of the pointer is $\xi(x)$, the wave function
after the post-selection becomes
\begin{equation}
	\label{pointer_wm}
	\xi^a_i(x)=(2\pi\delta^2)^{-1/4}e^{i\gamma\Im{\langle \boldsymbol{\sigma}^a_i\rangle_w}}\exp\left(-\frac{\left(x-\gamma\Re{\langle \boldsymbol{\sigma}^a_i\rangle_w}\right)^2}{4\delta^2}\right),
\end{equation}
for $\forall a\in\{0,1\}$.
Using Eq.~\eqref{pointer_wm}, the joint state of Alice's register,
Eve's memory, and Bob's pointer after the 
post-selection event (and tracing out Bob's qubit) is given by
\begin{equation}
	\rho_{AEP}^\prime=\frac{1}{2}\sum_{a\in\{0,1\}}\ketbra{a}_A\otimes\ketbra{\chi^a}_{EP}
\end{equation}
where
\begin{equation}
\label{chi_EP}
\ket{\chi^a}_{EP}=\sum_{i=1}^4\braket{\psi^a}{\Phi_i}\sqrt{\lambda_i}\ket{\nu_i}_E\otimes\ket{\xi_i^a}_P
\end{equation}
with $\ket{\xi_i^a}_P$ denoting the state of the pointer specified by wave function $\xi^a_i(x)$.
Bob then measures the pointer in the position basis.The state after this is described by

\begin{equation}
	\rho_{AEP}^{\prime\prime}=\frac{1}{2}\sum_{a\in\{0,1\}}\ketbra{a}_A\otimes\int_{-\infty}^{+\infty}P_a(x)\rho^{a}_{E}(x)\otimes\ketbra{x}dx.
\end{equation}
Here, normalized state $\rho^{a}_{E}(x)$
denotes Eve's memory corresponding to Alice's outcome $a$ when the pointer collapses to
position eigen state $\ket{x}$, and
\begin{equation}\label{P_a_x}
	P_a(x)=(2\pi\delta^2)^{-1/2}\exp\left(-\frac{\left(x-\gamma\Re{\langle \boldsymbol{\sigma}^a\rangle_w}\right)^2}{2\delta^2}\right)
\end{equation} 
denotes the probability of finding the pointer at position $x$ conditioned on the event that Alice gets outcome $a$,
where
\begin{equation}
	\langle \boldsymbol{\sigma}^a\rangle_w=\frac{\bra{\psi^a}\boldsymbol{\sigma}\rho_{AB}\ket{\psi^a}}{\bra{\psi^a}\rho_{AB}\ket{\psi^a}}
\end{equation}
is the weak value of $\boldsymbol{\sigma}$ for the pair prepared in mixed state $\rho_{AB}$ and post-selected in $\ket{\psi^a}$, given by Eq.~\eqref{gen_wv}.
Let $\tilde{P}(a,0)=P_a(\alpha)$ and $\tilde{P}(a,1)=P_a(-\alpha),
\forall a\in\{0,1\}$, and
\begin{equation}
	\tilde{P}=\sum_{a,b\in\{0,1\}}\tilde{P}(a,b).
\end{equation}
The ccq-state describing raw key registers of Alice and Bob,
and corresponding Eve's quantum memory,
given that Alice and Bob discard rounds when Bob gets outcome $\Pi_?$ in measurement 
$\mathcal{M}$, is expressed as

\begin{equation}
	\rho_{ABE}=\sum_{a,b\in\{0,1\}}P(a,b)\ketbra{a}_A\otimes\ketbra{b}_B\otimes\rho^{a,b}_{E}.
\end{equation}

Here $\ketbra{b}_B$ denotes the state of Bob's key bit when he gets outcome 
$\Pi_{b\in\{0,1\}}$. The joint probability distribution $P(a,b)$ is calculated as
$P(a,b)=\tilde{P}(a,b)/\tilde{P}$,
$\forall a,b\in\{0,1\}$. The state of Eve's memory conditioned on Alice's and Bob's
key bits reads
\begin{equation}
	\label{rho_ab_E}
	\rho^{a,b}_E=\rho^{a}_E((-1)^b\alpha), \forall a\in\{0,1\} 
\end{equation}
Note that $\Tr(\rho^{a,b}_E)=1$, $\forall a,b\in\{0,1\}$.

The correlation between the raw keys of Alice and Bob is quantified using the mutual information
$\mathcal{I}(A:B)$ with the joint probability distribution $P(a,b)$, and the mutual
information between Alice and Eve is upper bounded by the Holevo quantity
\begin{equation}
	\label{holevo}
	\chi(A:E)=S(\Omega_E)-\frac{1}{2}\left(S(\Omega^0_E)+S(\Omega^1_E)\right),
\end{equation}
where $S$ denotes von Neumann entropy, the state
\begin{equation}
	\Omega_{E}^a=\frac{P(a,0)\rho^{a,0}_{E}
		+P(a,1)\rho^{a,1}_{E}}{P(a,0)+P(a,1)}
\end{equation}
represents Eve's quantum memory corresponding to Alice's bit $a$,
and $\Omega_{E}=\left(\Omega_{E}^0+\Omega_{E}^1\right)/2$
is Eve's partial state. The secret key rate $r$ in asymptotic
limit with one-way  optimal error correction is lower bounded with
Devetak-Winter rate~\cite{Devetak2005},
\begin{equation}
	\label{DW}
	r\geq \ell_{DW}=\Omega\left[\mathcal{I}(A:B)-\chi(A:E)\right]
\end{equation}
where $\Omega$ is the post-selection probability. The protocol is secure when
$r>0$. The tolerable noise for secure protocol is then upper bounded by
\begin{equation}
	\eta_{tol}=\max\{\eta|\eta\in[0,1/2], \ell_{DW}>0\}.
\end{equation}

\section{Security analysis with weak measurement approximation}\label{sec_3.6}

Here derive the classical-classical-quantum (ccq) state of raw key bits
held by Alice and Bob, and the corresponding quantum memory of
Eve. Since we are only considering the asymptotic case under collective attack with
i.i.d. assumption, a mathematical description of only individual rounds is required
at the end for the security analysis. Moreover,
we evaluate expressions for the joint probability distribution of Alice and Bob
under the usual assumption of depolarizing quantum communication channel $\mathcal{E}(\cdot)$.
For the depolarizing channel, we have $\lambda_1=1-3\eta/2$,
and $\lambda_2=\lambda_3=\lambda_4=\eta/2$, where $\eta=\varepsilon_x=\varepsilon_y
=\varepsilon_z$ is the parameter quantifying the channel noise.
Therefore, $\rho_{AB}=\left(1-2\eta\right)\ketbra{\Phi_1}_{AB}+\frac{\eta}{2}\mathds{1}_{AB}$
and consequently, we have

\begin{equation}
	\begin{aligned}
		\langle\sigma^a\rangle_w
		&=\frac{\bra{\psi^a}\boldsymbol{\sigma}\rho_{AB}\ket{\psi^a}}{\bra{\psi^a}\rho_{AB}\ket{\psi^a}} \\
		&=\frac{(1-2\eta)\bra{\psi^a}\boldsymbol{\sigma}\ketbra{\Phi_1}\psi^a\rangle+\frac{\eta}{2}\bra{\psi^a}\boldsymbol{\sigma}\ket{\psi^a}}{(1-2\eta)\langle\psi^a\ketbra{\Phi_1}\psi^a\rangle+\frac{\eta}{2}}. \\
	\end{aligned}
\end{equation}
Using the facts that $\bra{\psi^a}\boldsymbol{\sigma}\ket{\psi^a}=0$, $\langle\psi^a\ketbra{\Phi_1}\psi^a\rangle=1/4$ for $a\in\{0,1\}$, and $\langle\boldsymbol{\sigma}^a_1\rangle_w=(-1)^a$, $\forall a\in\{0,1\}$, we get
\begin{equation}
	\label{WV_sigma}
	\langle\sigma^a\rangle_w
	=(-1)^a(1-2\eta).
\end{equation}
Therefore, the joint probability distributions of Alice and Bob becomes
\begin{equation}\label{joint_wma}
	P(a,b)=\left\{
	\begin{array}{rcl}
		\frac{1}{2\left(1+\exp(-\frac{2(1-2\eta)\gamma\alpha}{\delta^2})\right)} & \mbox{if} & a=b	\\ [0.5cm]
		\frac{1}{2\left(1+\exp(\frac{2(1-2\eta)\gamma\alpha}{\delta^2})\right)} & \mbox{if} & a\neq b,	\\
	\end{array}\right.
\end{equation}
and the raw key-bit error rate $Q=P(a\neq b)=P(0,1)+P(1,0)$, \ie~the probability that both parties generate
different key bits, is given by
\begin{equation}
	\label{QBER}
	Q = \frac{1}{\left(1+\exp(\frac{2(1-2\eta)\gamma\alpha}{\delta^2})\right)}.
\end{equation} 

\begin{figure*}[t] 
	\centering
	\begin{minipage}{0.48\textwidth}
		\centering
		\includegraphics[width=\linewidth]{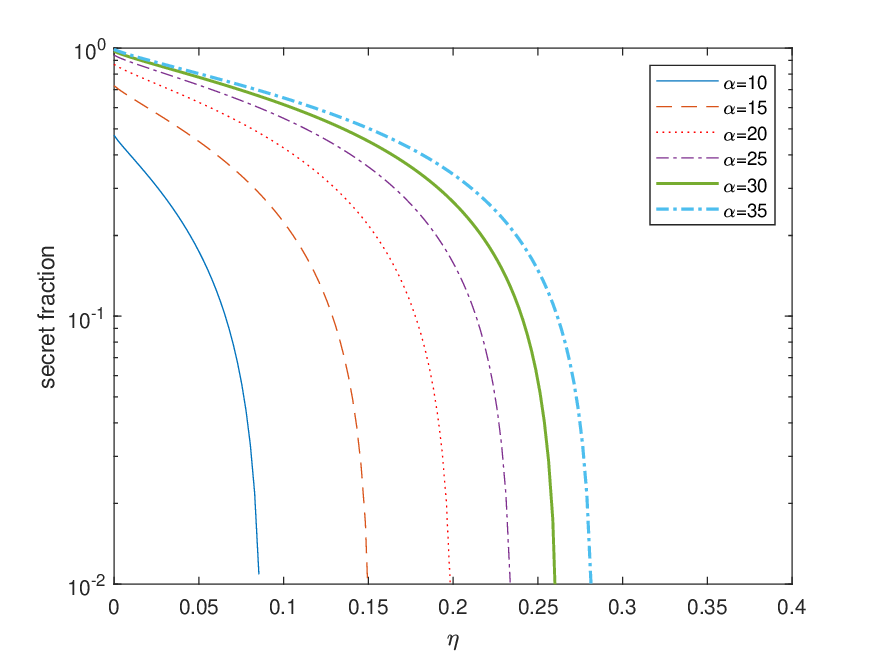} 
		\\[4pt] (a)
	\end{minipage}\hfill
	\begin{minipage}{0.48\textwidth}
		\centering
		\includegraphics[width=\linewidth]{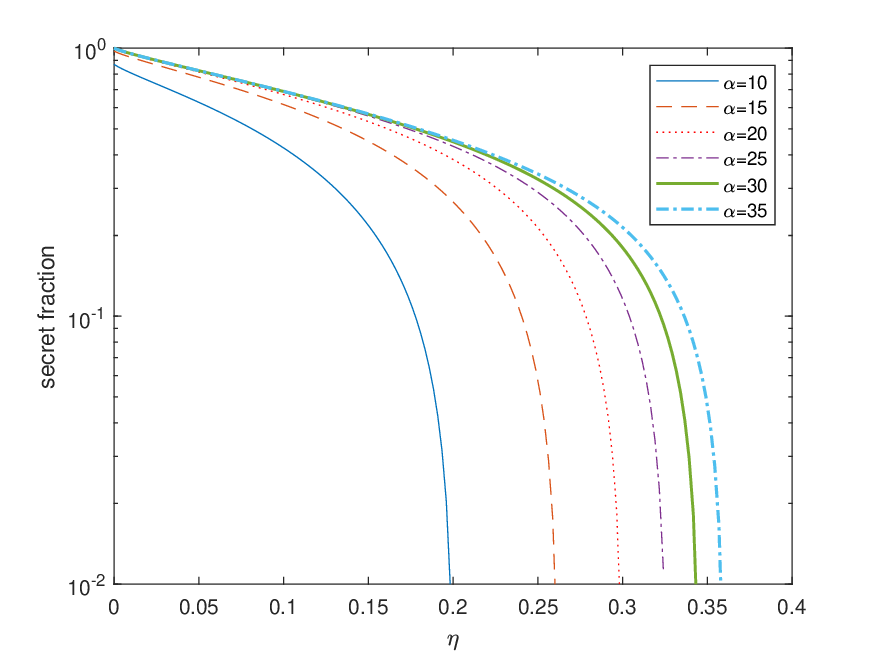} 
		\\[4pt] (b)
	\end{minipage}
		\caption{Secrete key fraction according to weak measurement approximation. The secret fraction is plotted as a function of depolarizing noise $\eta$ for (a) $\gamma=0.1$ and (b) $\gamma=0.2$.}
	\label{fig_sec_frac_wma}
\end{figure*}

The state of Eve's memory and Bob's pointer after the post-selection of shared qubit pair in $\ket{\psi^a}=\ket{a}\otimes\ket{+}$ is given by Eq.~\eqref{chi_EP} and, therefore, the state of Eve's memory $\rho^a_E(x)$ when the pointer collapses to $x$ is given as
\begin{equation} \label{eve_memory_x}
	\begin{aligned}
		\begin{array}{lcl}
			\rho^0_E(x)={\frac{1}{P_0(x)}
				\left(
				\begin{array}{cccc}
					\left(1-\frac{3\eta}{2}\right)\Vert\xi^+(x)\Vert^2   
					& \kappa\Vert\xi^+(x)\Vert^2 
					& \kappa\Vert\xi(x)\Vert^2
					& \kappa\Vert\xi(x)\Vert^2 \\ [0.4cm]
					\kappa\Vert\xi^+(x)\Vert^2 
					& \frac{\eta}{2}\Vert\xi^+(x)\Vert^2 
					& \frac{\eta}{2}\Vert\xi(x)\Vert^2
					& \frac{\eta}{2}\Vert\xi(x)\Vert^2 \\ [0.4cm]
					\kappa\Vert\xi(x)\Vert^2
					& \frac{\eta}{2}\Vert\xi(x)\Vert^2
					& \frac{\eta}{2}\Vert\xi^-(x)\Vert^2
					& \frac{\eta}{2}\Vert\xi^-(x)\Vert^2 \\ [0.4cm]
					\kappa\Vert\xi(x)\Vert^2
					& \frac{\eta}{2}\Vert\xi(x)\Vert^2
					& \frac{\eta}{2}\Vert\xi^-(x)\Vert^2
					& \frac{\eta}{2}\Vert\xi^-(x)\Vert^2 \\ [0.4cm]
				\end{array}
				\right)},	\\ [2cm]
			\rho^1_E(x)={\frac{1}{P_1(x)}
				\left(
				\begin{array}{cccc}
					\left(1-\frac{3\eta}{2}\right)\Vert\xi^-(x)\Vert^2   
					& -\kappa\Vert\xi^-(x)\Vert^2 
					& \kappa\Vert\xi(x)\Vert^2
					& -\kappa\Vert\xi(x)\Vert^2 \\ [0.4cm]
					-\kappa\Vert\xi^-(x)\Vert^2 
					& \frac{\eta}{2}\Vert\xi^-(x)\Vert^2 
					& -\frac{\eta}{2}\Vert\xi(x)\Vert^2
					& \frac{\eta}{2}\Vert\xi(x)\Vert^2 \\ [0.4cm]
					\kappa\Vert\xi(x)\Vert^2
					& -\frac{\eta}{2}\Vert\xi(x)\Vert^2
					& \frac{\eta}{2}\Vert\xi^+(x)\Vert^2
					& -\frac{\eta}{2}\Vert\xi^+(x)\Vert^2 \\ [0.4cm]
					-\kappa\Vert\xi(x)\Vert^2
					& \frac{\eta}{2}\Vert\xi(x)\Vert^2
					& -\frac{\eta}{2}\Vert\xi^+(x)\Vert^2
					& \frac{\eta}{2}\Vert\xi^+(x)\Vert^2 \\ [0.4cm]
				\end{array}
				\right)}.
		\end{array}
	\end{aligned}
\end{equation}
Here, we denote $\kappa=\sqrt{\left(1-\frac{3\eta}{2}\right)\frac{\eta}{2}}$, and 
\begin{equation}\label{xi_mp_wma}
	\xi^{\pm}(x)=(2\pi\delta^2)^{-1/4}\exp\left(-\frac{\left(x\mp\gamma\right)^2}{4\delta^2}\right)\\.
\end{equation}
Further, we assume $\xi^+(x)\xi^-(x)=\Vert\xi(x)\Vert^2\exp(-\gamma^2/2\delta^2)\approx\Vert\xi(x)\Vert^2$ according to the WMA. Eve's quantum memory conditioned on Alice and Bob's is computed using Eqs.~\eqref{rho_ab_E} and \eqref{eve_memory_x}.
We plot the Devetak-Winter secret fraction $F_{sec} = \mathcal{I}(A:B)-\chi(A:E)$ for different values of $\alpha$ and $\gamma$, see Figure~\ref{fig_sec_frac_wma}.
Surprisingly, introducing the QSD task using weak values and weak measurements improves the noise tolerance in the six-state QKD protocol. 
Surprisingly, introducing the QSD task using weak values and weak measurements improves the noise tolerance in the six-state QKD protocol. However, a careful investigation shows the contrary in the next section.



\section{Security analysis without weak measurement approximation} \label{sec_3.7}

In the previous section, we have calculated the secret fraction assuming weak measurement
approximation. Here, we re-analyze the security of the protocol without assuming
the weak measurement approximation ~\ie~retaining all powers of interaction
strength in calculations.

The state after applying $U_{BP}$ on $\ket{\Psi_{ABPE}}$
can be written without approximation as
\begin{equation}
	\begin{aligned}
		\ket{\Psi^\prime}&=U_{BP}\ket{\Psi}_{ABPE} \\
		&=\sum_{i=1}^4\sqrt{\lambda_i}U_{BP}\left(\ket{\Phi_i}_{AB}\otimes\ket{\xi}_P\right)\otimes\ket{\nu_i}_{E} \\
		&=\sum_{i=1}^4\sqrt{\lambda_i}\Big[\ket{\Phi_i}_{AB}\otimes\cos(\gamma\hat{p})\ket{\xi}_P
		-i\boldsymbol{\sigma}\ket{\Phi_i}_{AB}\otimes\sin(\gamma\hat{p})\ket{\xi}_P\Big]\otimes\ket{\nu_i}_{E} \\
	\end{aligned}
\end{equation}
The state of the pointer corresponding to Alice's bit $a$ and Bell state $\ket{\Phi_i}$
is expressed without approximation as
\begin{equation}
	\label{xi_i_a}
	\ket{\xi^a_i}_P=\exp(-i\langle\boldsymbol{\sigma}^a_i\rangle_w\gamma\hat{p})\ket{\xi}_P
\end{equation}
The joint probability distribution of Alice and Bob can be computed
using $P_a(x)$, which without approximation is given by
\begin{equation}
	\begin{aligned}\label{P_a1}
		P_a(x)
		&=\Tr{\Big(\ketbra{x}_P\otimes\mathds{1}_E\Big)\ketbra{\chi^a}_{PE}\Big(\ketbra{x}_P\otimes\mathds{1}_E\Big)^\dagger} \\
		&=4\sum_{i=1}^4{\lambda_i}\bra{\psi^a}\ket{\Phi_i}\bra{\Phi_i}\ket{\psi^a}\bra{x}\ket{\xi^a_i}\bra{\xi^a_i}\ket{x} \\
		&=\sum_{i=1}^4{\lambda_i}\Vert\xi^a_i(x) \Vert^2, \\
	\end{aligned}
\end{equation}
Let us now evaluate an expression for $\xi^a_i(x)=\braket{x}{\xi^a_i}$.
From Eq.~\eqref{xi_i_a}, we have
\begin{equation}
	\begin{aligned}
		\ket{\xi^a_i}
		&=\exp(-i\gamma\langle \boldsymbol{\sigma}^a_i\rangle_w\hat{p})\int_{-\infty}^{+\infty}\ket{x}\bra{x}\ket{\xi} dx \\
		&=\int_{-\infty}^{+\infty}\ket{x+\gamma{\langle \boldsymbol{\sigma}^a_i\rangle_w}}\bra{x}\ket{\xi} dx, \\
	\end{aligned}
\end{equation}
Since $\braket{x}{\xi}=\xi(x)=(2\pi\delta^2)^{-1/4}\exp\left(-x^2/{4\delta^2}\right)$,
we have
\begin{equation}
	\label{xi_i}
	\xi^a_i(x)=(2\pi\delta^2)^{-1/4}\exp\left(-\frac{\left(x-\gamma{\langle \boldsymbol{\sigma}^a_i\rangle_w}\right)^2}{4\delta^2}\right).	
\end{equation}
Eq.~\eqref{P_a1} can be re-written as
\begin{equation}
	P_a(x)=(2\pi\delta^2)^{-1/2}\sum_{i=1}^{4}\lambda_i\exp\left(-\frac{\left(x-\gamma{\langle \boldsymbol{\sigma}^a_i\rangle_w}\right)^2}{2\delta^2}\right).
\end{equation}
Since $\langle \boldsymbol{\sigma}^0_1\rangle_w=\langle \boldsymbol{\sigma}^0_2\rangle_w=\langle \boldsymbol{\sigma}^1_3\rangle_w=\langle \boldsymbol{\sigma}^1_4\rangle_w=1$
and $\langle \boldsymbol{\sigma}^1_1\rangle_w=\langle \boldsymbol{\sigma}^1_2\rangle_w=\langle \boldsymbol{\sigma}^0_3\rangle_w=\langle \boldsymbol{\sigma}^0_4\rangle_w=-1$, we have

\begin{equation} \label{P0P1}
	\begin{aligned}
		P_0(x)&=(2\pi\delta^2)^{-1/2}\left[(\lambda_1+\lambda_2)\exp(-\frac{(x-\gamma)^2}{2\delta^2}) +(\lambda_3+\lambda_4)\exp(-\frac{(x+\gamma)^2}{2\delta^2}) \right],\\
		P_1(x)&=(2\pi\delta^2)^{-1/2}\left[(\lambda_1+\lambda_2) \exp(-\frac{(x+\gamma)^2}{2\delta^2})+(\lambda_3+\lambda_4) \exp(-\frac{(x-\gamma)^2}{2\delta^2})\right].\\
	\end{aligned}
\end{equation}
Using Eq.~\eqref{xi_mp_wma}, Eq.~\eqref{P0P1} is re-written as
\begin{equation} \label{P0_P1}
	\begin{aligned}
		P_0(x)&=(\lambda_1+\lambda_2)\Vert\xi^+(x)\Vert^2+(\lambda_3+\lambda_4)\Vert\xi^-(x)\Vert^2,\\
		P_1(x)&=(\lambda_1+\lambda_2)\Vert\xi^-(x)\Vert^2+(\lambda_3+\lambda_4)\Vert\xi^+(x)\Vert^2.\\
	\end{aligned}
\end{equation}
Recall that we denote $\tilde{P}(a,b)=P_a((-1)^b\alpha)$. Using Eqs.~\eqref{P0_P1}, we can now express $\tilde{P}(a,b)$ as
\begin{equation}
	\begin{aligned}
		\tilde{P}(0,0)&=(\lambda_1+\lambda_2)\Vert\xi^+(\alpha)\Vert^2+(\lambda_3+\lambda_4)\Vert\xi^-(\alpha)\Vert^2,\\
		\tilde{P}(0,1)&=(\lambda_1+\lambda_2)\Vert\xi^+(-\alpha)\Vert^2+(\lambda_3+\lambda_4)\Vert\xi^-(-\alpha)\Vert^2,\\
		\tilde{P}(1,0)&=(\lambda_1+\lambda_2)\Vert\xi^-(\alpha)\Vert^2+(\lambda_3+\lambda_4)\Vert\xi^+(\alpha)\Vert^2,\\
		\tilde{P}(1,1)&=(\lambda_1+\lambda_2)\Vert\xi^-(-\alpha)\Vert^2+(\lambda_3+\lambda_4)\Vert\xi^+(-\alpha)\Vert^2.\\
	\end{aligned}
\end{equation}
For the case of depolarizing noise, we have
\begin{equation}
	\begin{aligned}
		\tilde{P}(0,0)&=\tilde{P}(1,1)=(1-\eta)P_+ + \eta P_-,\\
		\tilde{P}(0,1)&=\tilde{P}(1,0)=(1-\eta)P_- +\eta P_+,\\
	\end{aligned}
\end{equation}
where we denote
\begin{equation}\label{P_mp}
	P_{\pm}=(2\pi\delta^2)^{-1/2}\exp\left(-\frac{\left(\alpha\mp\gamma\right)^2}{2\delta^2}\right)\\.
\end{equation}
Note that, $P_-/P_+ = \exp(-2\gamma\alpha/\delta^2)$. Thus, using
\begin{equation}
	P(a,b)=\frac{\tilde{P}(a,b)}{\sum_{a,b\in\{0,1\}}\tilde{P}(a,b)},
\end{equation}
we can write the joint probability distributions of Alice and Bob as
\begin{equation}
	\label{joint_exact}
	P(a,b)=\left\{
	\begin{array}{rcl}
		\frac{(1-\eta)+\eta\exp(-\frac{2\gamma\alpha}{\delta^2})}{2\left(1+\exp(-\frac{2\gamma\alpha}{\delta^2})\right)} & \mbox{if} & a=b	\\ [0.5cm]
		\frac{(1-\eta)\exp(-\frac{2\gamma\alpha}{\delta^2})+\eta}{2\left(1+\exp(-\frac{2\gamma\alpha}{\delta^2})\right)} & \mbox{if} & a\neq b
	\end{array} \right.
\end{equation}
In order to calculate $\rho^{a,b}_E$, we first need to find $\rho^a_E(x)$ which is given by
\begin{equation}
	\label{28}
	\rho^a_E(x)=\frac{4}{P_a(x)}\sum_{i=1}^4\sum_{j=1}^4\sqrt{\lambda_i \lambda_j}\bra{\psi^a}\ket{\Phi_i}\bra{\Phi_j}\ket{\psi^a}{\bra{x}\ket{\xi^a_i}\bra{\xi^a_j}\ket{x}}\ketbra{\nu_i}{\nu_j}_{E}
\end{equation}

Note that $\bra{x}\ket{\xi^a_i}=\xi^+(x)$ if $\langle \boldsymbol{\sigma}^0_i\rangle_w=1$
and $\bra{x}\ket{\xi^a_i}=\xi^-(x)$ if $\langle \boldsymbol{\sigma}^0_i\rangle_w=-1$ for all $a\in\{0,1\}$ and $i\in\{1,2,3,4\}$.
Let us now denote $S^{\pm}=\Vert\xi^{\pm}(x)\Vert^2$, and
\begin{equation}
    S=\xi^+(x)\xi^-(x)=\Vert\xi(x)\Vert^2\exp(-\frac{\gamma^2}{2\delta^2}).
\end{equation}
The state $\rho^0_{E}(x)$ without weak measurement approximation can now be expressed in matrix form as
\begin{equation}
	\label{rho0_exact}
    \rho^0_{E}(x) = 
	\begin{aligned}
		\begin{array}{lcl}
			{\frac{1}{P_0(x)}
				\left(
				\begin{array}{cccc}
					\left(1-\frac{3\eta}{2}\right)S^+   
					& \sqrt{\left(1-\frac{3\eta}{2}\right)\frac{\eta}{2}} S^+
					& \sqrt{\left(1-\frac{3\eta}{2}\right)\frac{\eta}{2}} S
					& \sqrt{\left(1-\frac{3\eta}{2}\right)\frac{\eta}{2}} S \\ [0.2cm]
					\sqrt{\left(1-\frac{3\eta}{2}\right)\frac{\eta}{2}} S^+ 
					& \frac{\eta}{2}S^+
					& \frac{\eta}{2}S
					& \frac{\eta}{2}S \\ [0.2cm]
					\sqrt{\left(1-\frac{3\eta}{2}\right)\frac{\eta}{2}} S
					& \frac{\eta}{2}S
					& \frac{\eta}{2}S^-
					& \frac{\eta}{2}S^- \\ [0.2cm]
					\sqrt{\left(1-\frac{3\eta}{2}\right)\frac{\eta}{2}} S
					& \frac{\eta}{2}S
					& \frac{\eta}{2}S^-
					& \frac{\eta}{2}S^- \\ [0.2cm]
				\end{array}
				\right)},
		\end{array}
	\end{aligned}
\end{equation}
and, similarly, the state $\rho^1_{E}(x)$ can be expressed as
\begin{equation}
	\label{rho1_exact}
    \rho^1_{E}(x) = 
	\begin{aligned}
		\begin{array}{lcl}
			{\frac{1}{P_1(x)}
				\left(
				\begin{array}{cccc}
					\left(1-\frac{3\eta}{2}\right)S^-   
					& -\sqrt{\left(1-\frac{3\eta}{2}\right)\frac{\eta}{2}} S^-
					& \sqrt{\left(1-\frac{3\eta}{2}\right)\frac{\eta}{2}} S
					& -\sqrt{\left(1-\frac{3\eta}{2}\right)\frac{\eta}{2}} S \\ [0.2cm]
					-\sqrt{\left(1-\frac{3\eta}{2}\right)\frac{\eta}{2}} S^-
					& \frac{\eta}{2}S^-
					& -\frac{\eta}{2}S
					& \frac{\eta}{2}S \\ [0.2cm]
					\sqrt{\left(1-\frac{3\eta}{2}\right)\frac{\eta}{2}} S
					& -\frac{\eta}{2}S
					& \frac{\eta}{2}S^+
					& -\frac{\eta}{2}S^+ \\ [0.2cm]
					-\sqrt{\left(1-\frac{3\eta}{2}\right)\frac{\eta}{2}} S
					& \frac{\eta}{2}S
					& -\frac{\eta}{2}S^+
					& \frac{\eta}{2}S^+ \\ [0.2cm]
				\end{array}
				\right)}.
		\end{array}
	\end{aligned}
\end{equation}

\begin{figure*}[t] 
	\centering
	\begin{minipage}{0.48\textwidth}
		\centering
		\includegraphics[width=\linewidth]{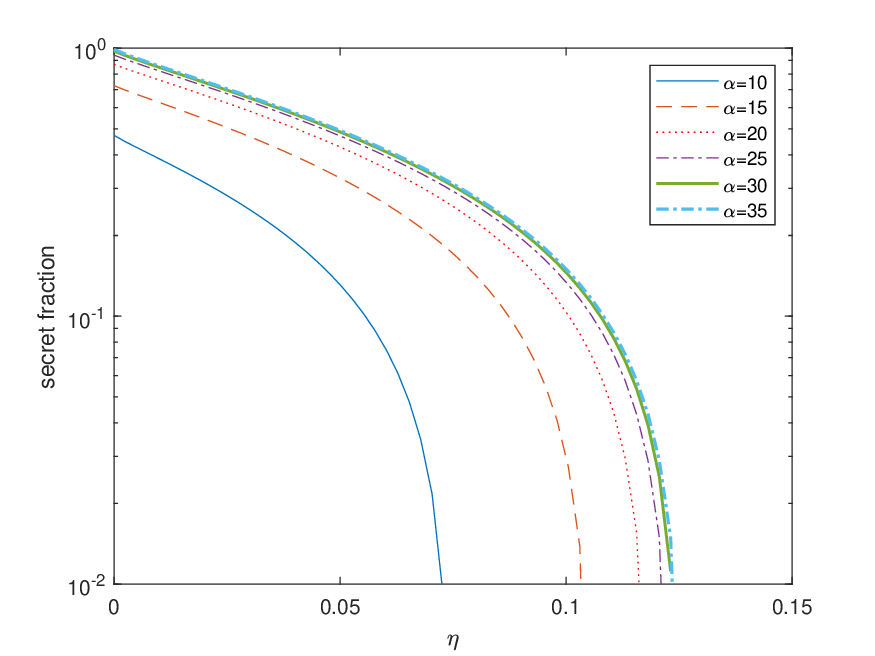} 
		\\[4pt] (a)
	\end{minipage}\hfill
	\begin{minipage}{0.48\textwidth}
		\centering
		\includegraphics[width=\linewidth]{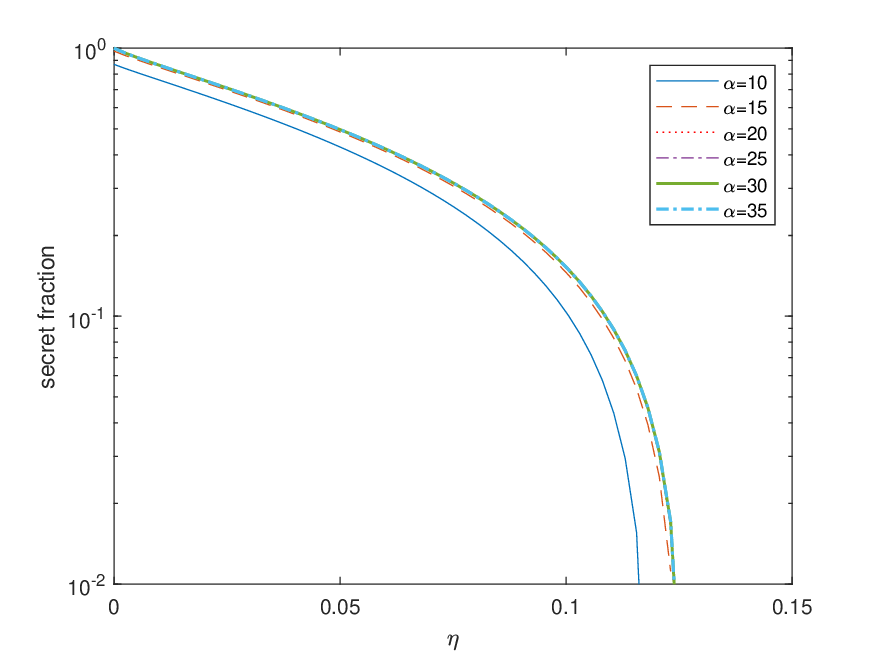} 
		\\[4pt] (b)
	\end{minipage}
	\caption{Secrete key fraction calculated without assuming the weak measurement approximation. The secret fraction is plotted as a function of depolarizing noise $\eta$ for (a) $\gamma=0.1$ and (b) $\gamma=0.2$, note that plots for $\alpha=20,25,30,35$ are coinciding.}
	\label{fig_sec_frac_exact}
\end{figure*}

Similarly to the case of weak measurement approximation,
the secret fraction $F_{sec}$ can now be computed using the joint probability given in
\eqref{joint_exact}, and Eve's memory states described by Eqs.~\eqref{rho0_exact}
and \eqref{rho1_exact}. In Figure~\ref{fig_sec_frac_exact}, we have plotted $F_{sec}$
for different values of $\alpha$ and $\gamma$. As it is clear from the plots, no positive
secret fraction was observed above the noise tolerance of the six-state protocol \ie~$12.62\%$.
In fact, for small $\alpha$ and $\gamma$, the secret fraction is smaller than
that of six-state protocol for the same noise. If we look carefully, the joint probability
distribution $P(a,b)$ in Eq.~\eqref{joint_exact} approaches the joint probability of the six-state
protocol as $\alpha$ is increased.
That means even with the use of a weak value-based state discrimination
scheme, the mutual information of Alice and Bob cannot exceed
what is observed in the six-state case. The latter is in contrast with what we saw in Section~\ref{sec_3.6}.

\section{Discussion and conclusions} \label{sec_3.8}

In this chapter, we have derived the weak value formalism for mixed states from the assumptions
of TSVF. Our generalization of weak values is the same as that proposed by other authors
who used different methods to formulate it~\cite{PhysRevA.65.032111,
	PhysRevA.89.012121,PhysRevLett.114.090403,PhysRevA.96.032114}.
We then devised a state discrimination scheme using weak measurements,
where we assumed the core properties of weak values and the weak measurement
approximation. Our scheme is motivated by the fact that two Gaussian distributions
can be distinguished with arbitrarily low error probability by selecting only out-layer events.
The formulation of weak values for mixed states was then used to discriminate mixed states in the six-state protocol. This approach apparently increased the noise tolerance drastically, giving an advantage over the original six-state QKD protocol. Moreover, this approach guarantees secure key generation at arbitrary high depolarizing noise.
However, we found that these exciting results are wrong and appear only because of
first order approximation in weak measurements. Moreover, these approximations are motivated
by TSVF and the assumption of weak values as elements of reality in weak measurements.
Our results have shown that such approximations must not be used without caution.
More interestingly, our quantum state-discrimination scheme may give the correct
answer for pure states but can fail in the case of mixed states. This puts a serious
caution on the uses and implications of generalized weak values. Contrary to what is implied
by TSVF (Section~\ref{sec_3.2}), weak values for mixed states might not be on equal footing
with those for pure states. We would also like to emphasize a direct implication of our
analysis that L. Vaidman's proposition that weak values are elements of the reality of weak
measurements~\cite{Vaidman1996,PhysRevA.96.032114} needs to be revisited and reanalyzed.

The author acknowledges the support of the Quant
Era grant “Quantum Coherence Activation By
Open Systems and Environments” QuCABOoSE 2023/05/Y/ST2/00139.


%

\end{document}